# Blockchain-Based Open Network in Technology Intermediation

1st Yang Yue
No. 2 Linggong Road, Ganjingzi District, Dalian City, Liaoning Province, P.R.C., 116024
Institute of Science Studies and S&T Management, Dalian University of Technology
Dalian, China
ysynemo@mail.edu.dlut.ed.cn

2nd Joseph Z. Shyu
No. 2 Linggong Road, Ganjingzi District, Dalian City, Liaoning Province, P.R.C., 116024
Institute of Science Studies and S&T Management, Dalian University of Technology
Dalian, China
joseph@cc.nctu.edu.tw

*Abstract*—Blockchain technology is developing using in reliable applications which can be designed to achieve decentralization and trustless. Based on the open network innovation theory, this paper proposes a technical intermediary management idea based on blockchain technology to improve the efficiency of technology intermediaries, providing accurate, reliable information and cutting cost for the market. This study demonstrates the advantage of blockchain to technology intermediaries. First, on a specific level, it can provide openness, transparency, decentralization and anonymity services. Second, the current industrial innovation elements are analyzed. blockchain improve the efficiency of technology intermediary, prevent risks and to make up for the shortcomings of traditional intermediaries. It has revolutionized the traditional technology intermediary. As this happens, it can revolutionize traditional technology intermediaries.

*Keywords—Blockchain, technology intermediary, open network, decentralization*

## I. INTRODUCTION

The blockchain technology has been booming in the industry, and its applications have become rapidly diversified. From the early private chain to multiple-centered distribution chain applications. Nakamoto (2008) published a paper called "Bitcoin: A Peer-to-Peer Electronic Cash System," and proposed the concept and algorithm of the electronic bitcoin currency, willing that it can build a system of "cryptographic-based rather than credit-based." The blockchain was initially defined as an anonymous trading system and is part of the Bitcoin protocol. The network relationship is to construct a peer-to-peer transaction mode transaction information, and the timestamped proof is available at each node —complete data storage. In this ecosystem, anyone can trade without knowing the other's background and need nothing intervention. The consensus mechanism is the core of the blockchain and there is no single management authority. For the very first time, China has stated its 13th Five-Year National Informaionization Plan for blockchain in 2016. Then in 2017, the Ministry of Industry and Information Technology released China's first blockchain standard "Blockchain Reference Architecture." The birth of the blockchain marks the beginning of humanity to build a genuinely credible Internet.

Technology intermediaries play an essential role for innovation. However, it has some problems , such as low efficiency but high costs, and low level of marketization and intractability. At present, the market needs a fair and transparent intermediary service to ensure these problem solved. The key point is to weaken the centrality status of technology intermediaries. The blockchain can make it, intermediary, more transparency and effectiveness.

Blockchain can be known as a revolution (or evolution) in institutions. Schwab (2017) predicted that the blockchain would become the key to a new wave of the industrial revolution. Blockchain changes the interaction mode, increasing system transparency and efficiency, especially in the shared chain of Bitcoin, where each user performs all the work that the intermediary does. It fosters a new generation of transactional applications that establish trust, accountability, and transparency. For the first time, NASDAQ deployed blockchain technology in its secondary market to secure the issuance and transfer of shares in private holding companies. OECD applied blockchain is a database that allows the transfer of value within computer networks. This technology is expected to disrupt several markets by ensuring trustworthy transactions without the necessity of a third party. With the potential value of the blockchains, UK government said blockchain is a type of database that takes several records and puts them in a block (rather like collating them on to a single sheet of paper). Each block is then 'chained' to the next block, using a cryptographic signature, which allows blockchains to use like a ledger, which can be shared and corroborated by anyone with the appropriate permissions.

This study uses the open network theory to study the blockchain technology intermediary to explore feasible solutions for technology intermediation based on blockchain. Analysis of the problems of technology intermediation and blockchain solutions under the open network theory, to make up for the shortcomings of the current research. This study aims to explore the integration of blockchain technology intermediaries, providing a new perspective for the practical application of technology intermediaries and blockchain technologies in the future. At the same time, it provides the government with a technical innovation proposal Research Method

## II. REASEARCH METHOD

This study uses qualitative research method and content analysis method. Qualitative research is used to review existing literature on technology intermediaries and blockchains. The content analysis method is a specific method for quantitative analysis of the content of the literature. Its purpose is to test the essential facts and trends in the novel, reveal the hidden information content contained in the research, and make intelligence predictions. At the same time, by using innovation intermediation model adapted from Henry, we illustrate the concepts of Blockchain-based technology intermediary platform in this study. We analyze every aspect of this platform, which can provide advantage services for technology intermediary. In industry level, the open innovation network consisting of the blockchain industry is used to construct a complete industrial value chain and based on the results of this study, blockchain and technology intermediation is perfectly matched.

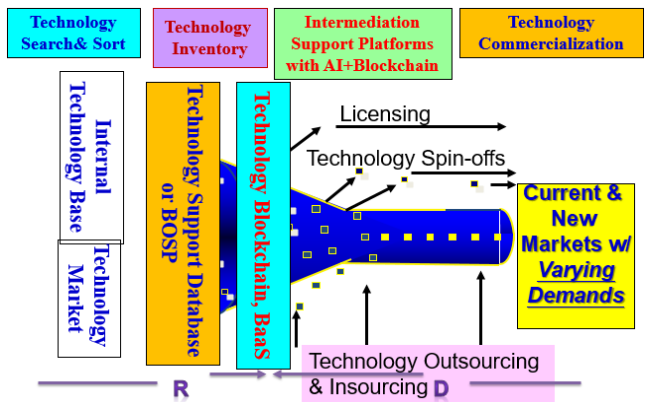

Fig. 1. Open innovation model of blockchain

Source: Adapted from Henry Chesbrough, Presentation to the Minnesota LES, May 18, 2006

## III. LITERATURE RIVIEW

### A. Blockchain

Blockchain has many advantages, such as anonymous transaction, decentralization, distributed storage, permanent storage, consensus mechanism, traceability, and non-tampering. According to the research purpose and research structure, we review and analyze the literature related to the blockchain and technology intermediation and review the related research for further discussion. There are many explanations about blockchain applications but little concern technology intermediation.

Davidsond (2016) Blockchain can be understood as a revolution in institutions, organization and governance. Mehrwald (2019) The combination of blockchain technology and smart contracts has the potential to facilitate non-intermediation and achieve actual peer-to-peer trading, provided that sufficient trust established and the conceptual model suggests the regulation and mediation of confidence in the online environment. Helps advance research on future experience trust related to blockchain technology in the shared economy. Malinova (2017) argues that implementing blockchain technology in financial markets can provide investors with new options for managing their shareholding transparency and their trading intentions. Covaci (2018) argues that blockchain is a distributed digital platform that is resistant to censorship and tampering, reducing many of the functions typically assumed by intermediaries while ensuring privacy, correctness and verifiability. Afanasyev (2019) presents an overview and classification of possible blockchain solutions for practical application of multiple generation systems and discusses how to use distributed ledgers to extend the research platform and libraries of existing multi-agent robot systems. Egelund(2017) express complex multi-party derivatives and conduct to automated execution. And he discusses how technology may change the way financial institutions, regulators, and individuals interact in a monetary system based on distributed ledgers. Swan (2018) said that the enterprise blockchain requires new business analysis and data science methods, such as deep learning algorithms and next-generation artificial intelligence solutions in the form of blockchains. Wang (2017)Construct a logistics model of the blockchain supply chain, the feasibility of applying blockchain technology to supply chain logistics from the main body, object dimension, trading mechanism, attribute aspect, original contract and functional size. Internal demand shifts to external linkages to reduce market uncertainty. Casado (2019) describes a new blockchain-based architecture for creating e-health systems. Transactions between blockchains and wireless sensor networks are conducted through smart contracts, without the need for intermediaries and preventing human error.

Sharples(2016) undertaking initial trials of a private blockchain or storing educational records, drawing also on our previous research into reputation management for educational systems. Yang(2018) build a dynamic and flexible financial regulatory system could effectively address Fintech risks. Technology-driven regulations focused on data monitoring could be a remedy for the inefficiency and ineffectiveness of traditional commercial rules and enhance adequate protection of financial consumers' rights and interests. Skowronski(2019) discuss an intrinsic communication protocol capable of rewarding intermediaries of data exchange for each transmitted byte and tackle the feasibility, security and accountability challenges which we have faced and which will need to be considered by future architects of the new, exciting family of cyber-physical systems. Van(2019)states current shareholder engagement systems face massive classical inefficiencies. First, due to the large chains of intermediaries in the current securities models, transaction costs are high, and shareholder votes and other information are not always correctly transmitted between shareholders and issuers. With permission, data stored in a correct and immutable way, with a consensus mechanism tailored to its purpose. Hasan(2018) shows an arbitration mechanism is also incorporated if a dispute arises during the shipping process. In this paper, and how they implemented, verified, and tested the proper functionality of our PoD solution. In practice, Ramezan propose a novel blockchain-based contractual routing (BCR) protocol for a network of untrusted IoT devices, and the result shows that the BCR protocol enables distributed routing in heterogeneous IoT networks. Zamani(2018) argue for the role of the blockchain distributed ledger technology, in building innovative business models, including machine money, autonomous economic agents and decentralized organizations. Findings The authors highlight three application areas for blockchains, whereby they can function as applications, can help develop independent commercial agents and can lead the development of decentralized autonomous organizations.

With regards to the question of market disintermediation, the authors suggest that, rather than complete disintermediation, the most probable scenario is that of new types of intermediaries finding previously unthinkable roles to play in mediating blockchain-based economic transactions.

Although the mainstream view tends to be a blockchain that can revolutionize the intermediary, there are also some contrary views. Pavel Kravchenko believes that for government, voting platforms, trading platforms and other application scenarios, verifying identity and entering data is required. The role of the role still requires a specific purpose and cannot be borne by the individual. Primavera de Filippi believes that the problem of blockchain is not the trust algorithm, but the person who is engaged in the technology, pointing out that the trust in the blockchain is moving from some human intermediaries to the algorithm and then back to the technicians themselves.

*B. Technology intermediation*

The technology intermediary is a third-party identity market participant who plays an essential role in entering the market, participating in the competition, maintaining market order, and handling market disputes. It provides services such as assessment, coordination, impartiality, arbitration, and professional institutions that offer exchanges, agents, and consulting services. Technology intermediaries play a role in solving information asymmetry, reducing information search costs, efficiently matching efficiency, reducing transaction costs, and controlling transaction risks. The study of technology intermediaries is various. Howells(2006) mentioned the concept of technology intermediaries, which considered that technology intermediaries refer to institutions that provide services for the social transfer of technology. The commercial application of technological achievements and the social diffusion of knowledge. At the same time, the study of technology intermediaries summarized into Technology transfer, innovation management, technology systems, intermediary organizations. From the perspective of technical methods, technology intermediaries can be used as extensions of R&D institutions and complement the functions of the technology market, providing services for functional activities, providing professional services throughout the organization from the production to the transfer of the application.

Intermediation Services is a Service for Manufacturing, entailing innovation-intensive technology intermediation and open innovation, technology customization and specialization, globalized management in technology integration, and professional consulting and logistics services. The intermediary defined as a system that complements organizational categories, improving, transforming, and ensuring that systems function by reducing the complexity of transactions. While technology intermediaries build bridges outside the market, they use open innovation to promote the transfer of new technical knowledge and are an essential part of national or regional innovation systems. Benthon believes that there will be no intermediary in a market where there is no friction in transaction costs, information costs and indivisibility. Fang asserts that the information asymmetry in the technology market causes the problem of adverse selection in technology transactions. Serious adverse selection hinders the transformation of technological achievements and reduces the transaction efficiency of the technology market. Chen believes that the intermediary charges the service fee between the buyer and the seller and the three factors related to the transaction cost saved by the two parties in choosing the intermediary transaction mode, the value of the intermediary enterprise itself, and the risk cost faced by the buyer and the seller. Rothwell and Zegveld (1981) summarize the nine resource factors needed for industrial innovation, including the technical knowledge and human resources, market information and management skills, financial resources, research and development, research environment, domestic market, foreign market, internal market environment, and international market environment, etc. Leyden (1992) propose that innovation leads to technology, and technology is the primary driver of economic growth. Shyu (1999) suggests focusing on how to guide the modernization of the industry to change the constraints and create new competitive advantages.

Gindy (2006) suggests the successful acquisition and management of technology to develop and manufacture innovative products is a critical factor in their competitiveness. According to Howells' interpretation of technology intermediaries, it is believed to have emerged as an intermediary role in the agricultural trading process of the 16th century. These people not only participate in transactions but also actively disseminate agricultural technological improvements as informal communicators. It considered that the technical intermediary acts as a broker or a third party, and is a bridge in the open system. Connectivity is an essential participant in the innovation system. After continuous research by scholars, technology intermediaries currently involve four levels of study, including technology transfer and transfer, innovation management, technology systems, and intermediary organizations. Jakob(2016) Innovative intermediaries help to establish or improve connections between different actors based on reciprocal relationships of skills or interests to support the generation and dissemination of innovation. The technical intermediary is a professional institution that enters the market as a third party and provides services such as agency, consulting, etc. for evaluation, coordination, and arbitration. Technology intermediaries play a role in solving information asymmetry, reducing information search costs, efficiently matching efficiency, reducing transaction costs, and controlling transaction risks. Mooney (1990) believes that intermediation solvency and integrity are the core of processing, and the integrity of intermediaries is used to improve customer asset security in the intermediary system. Colomo (2010) The management of mediation is a very complicated process. In this scenario, a reliable information technology platform is needed to manage dynamic information. From the perspective of technical systems, technology intermediaries can be used as extensions of R&D institutions and complement the functions of the technology market. FUKUGAWA (2016) Study of the Japan Local Public Technology Center (LPTC), which arranges various technology transfer channels for small and medium-sized enterprises in the region, and solves multiple (technical and non-technical) problems through professional consultation under design. When it is challenging to explain, LPTC acts as an intermediary for innovation, linking SMEs to other sources of knowledge such as universities. Domestic scholars' research on technology intermediaries is more from the perspective of technology intermediary organizations and technology transfer. Fang(2003) believes that the information asymmetry in the technology market causes the problem of adverse selection in technology transactions. Serious adverse selection hinders the transformation of technological achievements and reduces the transaction efficiency of the technology market. Chen(2004) believes that the intermediary balance the pricing of buyers and sellers is affected by three aspects, including the transaction cost saved by the two parties in choosing the intermediary transaction mode, the value of the intermediary enterprise itself, and the risk cost faced by the buyer and the seller

IV. TECHNOLOGY INTERMEDIATION BLOCKCHAIN

A. *Intermediary via blockchain*

Technical intermediaries establish bridge relationships outside the market. However, there are many problems with technology intermediaries. Firstly, technology R&D institutions and personnel rarely can allocate market resources, especially the ability to promote technology in the market. Second, technical transaction entities need to be Identify, and they can finally complete technology transfer through negotiation. Third, the problem of middleman trust widely exists in the intermediary. Reputation is the central character of intermediary. Forth, Lack of cross-regional, interdisciplinary, networked collaborative models, etc. In this process, technology intermediaries need to be more comprehensive.

Blockchain technology is a reliable, traceability, trustworthy database technology which can be designed to achieve decentralization, trustless, and distributed ledger applications. The blockchain can improve the efficiency of technology intermediaries, which can solve the problem of low capability of technology intermediaries. Blockchain technology has revolutionized technology intermediation through decentralization and weak centralization mechanisms. It provides accurate and reliable technology and market information between supply and demand sides, saving the cost of technology transfer processing, that can provide (table 1) openness, autonomy, peer-to-peer, decentralization, autonomy, security, reliability, no tampering, transactional anonymity, transparency and other excellent features.

TABLE I. TECHNOLOGY INTERMEDIATION VIA BLOCKCHAIN

| | | **Blockchain** | **Intermediation Failure** | **Blockchain Advanced** |
|---|---|---|---|---|
| Decentralization | Openness | (1) Anyone verified can join the system. (2) Everyone can query the blockchain data and develop related applications through the public interface. | (1) Distinct regional fragmentation. (2) Low degree of marketization. (3) Industry barriers | (1) The efficiency of technology intermediation in the open system is higher. (2) Eliminate the multi-party information asymmetry and unclear background in the technology intermediary. |
| | Decentralize | Distributed storage and computing, node equal rights. | High cost of establishing and operating a reliable intermediary and lower security under cyber attack | Decentralization or weakness centralization, reducing the intermediary risk and the price |
| | Autonomy | Implementing a programmable society through smart contracts, with which all nodes exchange data freely and securely in a trusted environment. | (1) High management costs. (2) Low efficiency. (3) High labor costs. (4) Human error cannot be avoided. | (1) Reduce time costs: Smart contracts greatly reduce time costs. (2) Write automated transactions and related policies into smart contracts, so that there is no human intervention in the entire process. (3) The process is transparent and the results are accurate. (4) The user's trust is improved, at the same time to reduce contract disputes. |
| | Trustless | A peer-to-peer network consisting of many nodes. Data exchange between nodes relies on a consensus mechanism rather than trust. | Middleman trust problem: Reputation is the crucial character of intermediary. | |
| | P2P | End-to-end network | The lack of networked coordination of technical intermediaries. | Ensure that the supply and demand sides are precisely matched. |
| Security | Security | Distributed storage and computing to make blockchain network data reliable. | The disadvantage of technology cumulative | Blockchain enables efficient access in cloud computing environments, ensures storage and computational efficiency, improves blockchain query efficiency, improves system stability and robustness, and improves data security and data reliability. |
| | Continuity | Once the data is verified, the data and timestamp are stored forever. | Insufficient data | Using blockchain to store user data, can effectively take advantage of the inherent advantages of technology itself, and strengthen risk assessment, etc. |
| | Unchangeable | Only can be modified when the computing power exceeds 51% | Unknown internal risk control capability | (1) Improve the internal risk control capability of technology intermediaries,<br>(2) Ensure the security of the ledger system, funds and information, establish a blockchain general ledger system, improve financial security |
| | Anonymous | Identity and other people's transaction history is anonymous | The need to provide complex personal information in the process of trading is not only cumbersome, but also the risk of information lick. | User privacy and trade secrets is protected, and the content of the contract is only accessible to the parties. |
| Traceability | Traceability | The information, rules is highly transparent: All the data, trading Information, time are all openness. Every transaction is visible at all nodes | Technical intermediary audit problems | (1) Audit work can be done online through a granted auditor node.<br>(2) All regulatory records will also be saved to ensure just. |

*B. Blockchain-based intermediary system*

Constructing block technology intermediary support information system for cost control: Through technology intermediary information system, it can realize data security audit, record, intellectual property transaction and management, property deed, legal application and other functions. Blockchain BaaS enhances BOSP's secure, trusted, and anonymous resolution of tagging, data tracking, data cross-referencing, big data, data sharing, and related research and management data issues.

Develop an intelligent Energy Management Platform, focusing on Power Management Systems Services and Technology Intermediation, iEMS effectively and efficiently develops a Service platform which facilitates dynamic customer interactions, and delivers individualized systems applications. The Key Features of iEMS Platform, such as outage management, distribution management, condition-based maintenance, supervisory control & Data acquisition (SCADA), distribution planning, Load & Demand Forecasting, advanced metering infrastructure and policy expertise.

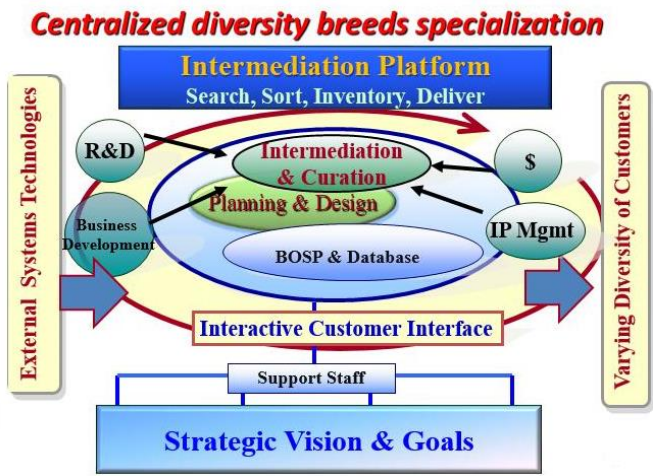

Fig. 2. centralized diversity breeds specialization

Source: Joseph Z. Shyu

Technology Intermediaries, which involves complex legal environments and large-scale data interactions, implements trusted transactions of data through blockchain technology, breaks existing barriers to profit, and creates a new ecosystem of internal and external security sharing in the data industry. Blockchain technology intermediation is not only a technology platform or business process platform, but also an automation community with top-level design, providing accurate and reliable technology and market information, building information support systems, accelerating the transformation of industrial technology, and integrating markets. Resources, technology foresight and diagnosis, knowledge processing and knowledge reorganization, gatekeepers and intermediaries, testing and verification, identification, verification, management, protection, evaluation and commercialization of technical results..

V. FINDINGS AND DISCUSSION

This study intends to analyse blockchain intermediation from the industry level to explore the future development of blockchain technology. Through the two industrial level profiles of blockchain and technology intermediation system, the current industrial innovation elements are analysed. The blockchain-based technology intermediary platform provides accurate and reliable technical support. it can provide professional intermediary services. Therefore, technology intermediaries make technology transfer, activities procedural, reduce transaction costs, improve transparency, improve the efficiency of market information exchange and risk prevention, which has completely changed the traditional technology intermediary. It can also serve as knowledge of intelligence units and knowledge bases. Process and collect information, diagnose, and identify and monitor relevant technologies and expertise in the technology market in a gatekeeper or facilitation manner. Identify and disseminate new and valuable knowledge and technology. In such deploying a technology library, you can enhance the ability to transform technology.

*1) The technology intermediation in the open network system is more effciency:*

*a)* Smart contracts greatly reduce time costs..

*b)* Write automated transactions and related policies into smart contracts, so that there is no human intervention in the entire process.

   *2) The process is transparent and the results are accurate:*

    *a)* All regulatory records can be saved clearly and easy to use.

    *b)* Audit work can be done online through a granted auditor node.

    *c)* The user's trust is improved, at the same time to reduce contract disputes. A credit society can be founded through smart contract, increasing credit between users and thereby reducing contract disputes during the process of transaction.`

    *d)* Eliminate the information asymmetry problem and suspicion reputation of technology intermediary.

   *3) Very important deal is to achieve a decentralization or weakness centralization ecosystem.*

    *a)* Improve the internal risk control capability of technology intermediaries..

    *b)* Ensure the security of the ledger system, funds and information, establish a blockchain general ledger system, improve financial security.

   *4) Security may be the most typical feature of blockchain:* Blockchain enables efficient access in cloud computing environments, ensures storage and computational efficiency, improves blockchain query efficiency, improves system stability and robustness, and improves data security and data reliability. Using blockchain to store user data, can effectively take advantage of the inherent advantages of technology itself, and strengthen risk assessment, etc.

## VI. PRACTICAL IMPLICATIONS AND LIMITATIONS

In this study, Blockchain-based technology intermediation was discussed and combined with relevant literature and methods. Through analyzing two separate industrial level profiles of blockchain and technology intermediation, the current industrial innovation elements of technology intermediation are analyzed. Blockchain-based technology intermediary provide accurate and reliable technology support and market information exchanging, to improve the efficiency of technology intermediary, prevent risks and to make up for the shortcomings of traditional technology intermediaries. It has revolutionized the traditional technology intermediary. However, the landing of the blockchain technology intermediary still have long way to go.

- The consistency of information under the chain is difficult to guarantee. The regulatory problems caused by the inability to modify the uplink data require rigorous identity and data verification. The internal data of the blockchain is dependent on the real world. every information up to blockchain is by human. The interaction between data inside and outside the blockchain cannot avoid the intermediaries, the human. So, the digital ownership platform requires intermediaries to verify your identity. the blockchain avoids the existence of intermediaries, now the truth is the data on blochchian is insufficient..

- The stability of the blockchain technology itself needs to be improved. It is not mature enough in the application security and business model of the blockchain. It is necessary to optimise the underlying architecture and actively cultivate compound talents continuously. Build a professional R & D team..

- The achievement of industry consensus still requires a process, while lacking a unified standard for the blockchain. In order to achieve industrialisation and form an ecological industrial chain, standardisation is the only way to go. The blockchain technology has not yet built a unified standard at home and abroad, and gradually develops standards related to the blockchain, which will help to realise the landing of the blockchain technology intermediary platform finally.

- The legal system of blockchain technology intermediation needs to be improved. No matter how well the blockchain technology is conducive to social progress, the role behind the technology is still human. Computer code is increasingly used to regulate our online behaviour to replace laws and regulations, while legal regulations lag behind technological advances. Illegal transactions in the blockchain are also protected by anonymity. Lack of control will undoubtedly undermine the development of technology intermediaries.

However, as mentioned above, further research is needed. A blockchain-based technology intermediation further needs to be studied from smart contracts and policy support. the smart contract is the key to how the system's automatic operation well. Blockchain, as one of the most motivating technologies, encourage us to look forward to building a foundation for early research to support this new field.